\documentclass{article}

\usepackage{amssymb}
\usepackage{amsthm}
\usepackage{amsmath}
\usepackage{graphicx}
\usepackage{subfigure}
\usepackage[figuresright]{rotating}
\usepackage{amsfonts}
\usepackage{latexsym}
\usepackage{bbm}
\usepackage{mathrsfs}
\usepackage{varioref}
\usepackage{hyperref}

 \graphicspath{{figures/}}

\newcommand{\widesim}[2][1.5]{
  \mathrel{\overset{#2}{\scalebox{#1}[1]{$\sim$}}}
}

\title{Gridless particle technique for the Vlasov-Poisson system in problems with high degree of symmetry}
\usepackage{authblk}

\author[1]{E. Boella}
\author[2]{G. Coppa}
\author[3]{A. D'Angola}
\author[4]{B. Peiretti Paradisi}

\affil[1]{Center for Mathematical Plasma Astrophysics, KU Leuven, Leuven, Belgium}
\affil[2]{Dipartimento di Elettronica e Telecomunicazioni, Politecnico di Torino, Torino, Italy}
\affil[3]{Scuola di Ingegneria, Universit\`a della Basilicata,  Potenza, Italy}
\affil[4]{Dipartimento Energia, Politecnico di Torino, Torino, Italy}

\begin{document}
	\maketitle
\begin{abstract}
In the paper, gridless particle techniques are presented in order to solve problems involving electrostatic, collisionless plasmas. The method makes use of computational particles having the shape of spherical shells or of rings, and can be used to study cases in which the plasma has spherical or axial symmetry, respectively. As a computational grid is absent, the technique is particularly suitable when the plasma occupies a rapidly changing space region.
\end{abstract}

\section{Introduction} \label{intro}
The work investigates the possibility of using gridless particle
techniques \cite{Dawson-PoF-1962, Eldridge-PoF-1962} in the study of
plasmas which are produced by laser-matter interaction with the
purpose of accelerating positive ions. Avoiding to introduce a
computational grid is useful in situations (as for plasma expansions
and explosions), in which the physical domain occupied by the
particles increases rapidly in time. In this framework, in general
situations one could employ a set of computational particles and
directly calculate the electric field acting on each of them, as the
sum of the contribution of the other particles. This requires an
extremely high computational effort, unless the problem under exam
presents some symmetry. In the work, the cases of spherical and
axial symmetry are considered. In the first case (Sect. 2), the
problem is essentially one dimensional and computational particles
are in the shape of spherical shells. By using the Gauss's formula,
the electric field is readily evaluated. For the second case (Sect.
3), particles are modeled as thin circular rings, which are
characterized by their radii and their axial coordinates. In this
case, the evolution of the force acting on each particle requires
necessarily the calculation of the sum of contributions due to the
other particles. Although some advantages which are present in the
spherical case are lost, the technique here presented conserves
interesting features also in this case. Results for both cases are
shown and they are compared with exact calculations (when available)
or with Particle-In-Cell simulations.

\section{The shell method} \label{shell}
This Section presents in a complete, rigorous way the method of the shells, which was already introduced and employed with different formulations by other Authors (in particular, in refs. \cite{Dawson-PoF-1962, Eldridge-PoF-1962, russi}). 
\subsection{First formulation} \label{ff}
In its simplest formulation, a set of $N$ computational particles is considered. After initializing their coordinates $\mathbf{x}_i$
and momenta $\mathbf{p}_i$, the particle are ordered according to their radial coordinates $r_i=|\mathbf{x}_i|$, so that $r_j>r_i$ if $j>i$. Then the radial electric field acting on each particle is evaluated simply as:
\begin{equation}
  \mathbf{E}_i=\left( \sum_{j=1}^{i-1}  q_j+ \tfrac{1}{2}q_i \right)  \frac{\mathbf{x}_i}{r_i^3},
\label{electric_field}
\end{equation}
by using the Gauss's formula and taking advantage of the spherical
symmetry of the problem. The presence of the factor $\tfrac{1}{2}$
multiplying $q_i$ can be explained in a simple way by considering
that, for $r=r_i-\epsilon$ ($\epsilon \rightarrow 0^+$) $q_i$ does not contribute
to the electric field, while for $r=r_i+\epsilon$ the total charge
to be evaluated is $\sum_{j=1}^{i}  q_j$. Thus, by supposing a
linear behavior of $\mathbf{E}$ at the interface, the factor
$\tfrac{1}{2}$ provides the correct value of the field (a rigorous
proof of the formula is presented in Sect. 2.4). Finally, after
evaluating $\mathbf{E}$ on each computational particle, the
equations of motion:
\begin{equation}
\begin{cases}
 \frac{\displaystyle \mathrm{d} \mathbf{x}_i} {\displaystyle \mathrm{d}t}= \frac{\displaystyle \mathbf{p}_i}{\displaystyle m_i}, \\
  \frac{\displaystyle \mathrm{d} \mathbf{p}_i}{\displaystyle \mathrm{d}t} = q_i\mathbf{E}_i(\mathbf{x}_1,\mathbf{x}_2,...,\mathbf{x}_N)\label{part_push},
\end{cases}
\end{equation}
can be solved by using a suitable numerical technique (e.g., the leapfrog or the Runge-Kutta method), using a time step much smaller with respect to the inverse of the plasma frequency.

\subsection{Second formulation} \label{sf}
The technique described above is very simple (for example, a MATLAB code can be implemented in few lines of program), but it is excessively
memory and time consuming, as it does not take fully advantage of the symmetry of the problem. In fact, in a central field of forces, the
trajectory of each particle takes place on a plane. Therefore, the motion is essentially a two-dimensional problem. This fact suggests a
new, simpler formulation of the method. After generating the initial 3D coordinates $\mathbf{x}_i$ and momenta $\mathbf{p}_i$, a set of 2D coordinates $\mathbf{X}_i$ and $\mathbf{P}_i$ is defined as
\begin{equation}
\begin{cases}
       \mathbf{X}_i = (r_i,0), \hspace{1cm}  i=1,2,...,N,\\
       \mathbf{P}_i = \left(\mathbf{p}_i  \cdot \frac{\displaystyle \mathbf{x}_i}{\displaystyle r_i} , \left|\mathbf{p}_i - \left(\mathbf{p}_i  \cdot \frac{\displaystyle \mathbf{x}_i}{\displaystyle r_i}\right)\frac{\displaystyle \mathbf{x}_i}{\displaystyle r_i}\right| \right).
 \label{part_push}
\end{cases}
\end{equation}
After that, the method is completely identical to the previous formulation, but it uses only 2D vectors. More in detail, the particles are ordered according to the radial position $R_i=|\mathbf{X}_i|$, the electric field is evaluated as
\begin{equation}
    \mathbf{E}_i=\left( \sum_{j=1}^{i-1}  q_j+\tfrac{1}{2}q_i \right) \frac{\mathbf{X}_i}{R_i^3}, \label{electric_field2d}
\end{equation}
and the evolution of the system is governed by the equations
\begin{equation}
\begin{cases}
 \frac{\displaystyle \mathrm{d} \mathbf{X}_i} {\displaystyle \mathrm{d}t}= \frac{\displaystyle \mathbf{P}_i}{\displaystyle m_i},  \\
  \frac{\displaystyle \mathrm{d} \mathbf{P}_i}{\displaystyle \mathrm{d}t} = q_i\mathbf{E}_i(\mathbf{X}_1,\mathbf{X}_2,...,\mathbf{X}_N).
\label{pp2D}
\end{cases}
\end{equation}

\subsection{Third formulation} \label{tf}
Starting form the Lagrangian
\begin{equation}
\mathscr{L}\left(r,\varphi,\dot{r},\dot{\varphi},t\right)=\dfrac{m}{2}\left(\dot{r}^{2}+r^{2}\dot{\varphi}^{2}\right)-q\Phi\left(  r,t\right),
\label{lagcic}
\end{equation}
for a single particle in a central potential ($\Phi$ depends on $t$ due to the interaction with the other particles of the plasma), one can obtain the Hamiltonian
\begin{equation}
\mathscr{H}\left(r,\varphi,p_{r},p_{\varphi},t\right)=\frac{1}{2m}\left(p_{r}^{2}+\frac{p^2_{\varphi}}{r^{2}}\right)+q\Phi\left(  r,t\right),
\label{hcic}
\end{equation}
and the equations of the motion
\begin{equation}
\begin{cases}
\frac{\displaystyle \mathrm{d} r} {\displaystyle \mathrm{d}t}= \frac{\displaystyle p_r}{\displaystyle m}, \hspace{3cm}  \frac{\displaystyle \mathrm{d} \varphi} {\displaystyle \mathrm{d}t}= \frac{\displaystyle p_{\varphi}}{\displaystyle mr^2},  \\
 \frac{\displaystyle \mathrm{d} p_r} {\displaystyle \mathrm{d}t}= -q\frac{\displaystyle \partial\Phi}{\displaystyle \partial r}+\frac{\displaystyle p^2_{\varphi}}{\displaystyle mr^3}, \hspace{1cm} \frac{\displaystyle \mathrm{d} p_{\varphi}} {\displaystyle \mathrm{d}t}= 0.
\end{cases}
\end{equation}
In other terms, as it is well known, for a central potential there is a constant of the motion, $p_{\varphi}$, which corresponds to the axial angular momentum,
and the motion in radial direction is essentially one-dimensional. This suggests a third way of studying the dynamics of these systems. Starting again from the set $\{ \mathbf{x}_i, \mathbf{p}_i \}$ one can calculate
\begin{equation}
r_i=|\mathbf{x}_i|, \hspace{1cm} p_{r,i}=\mathbf{p}_i  \cdot \frac{\displaystyle \mathbf{x}_i}{\displaystyle r_i}, \hspace{1cm} p_{\varphi,i}=r_i \left|\mathbf{p}_i - p_{r,i}\frac{\mathbf{x}_i}{r_i}\right|.
 \label{eq:prz}
\end{equation}
Then, the radial electric field is evaluated as
\begin{equation}
    {E}_{r,i}=\left( \sum_{j=1}^{i-1}  q_j+\tfrac{\displaystyle 1}{\displaystyle 2}q_i \right) \frac{1}{r_i^2} \label{efield2d}
\end{equation}
(of course, particles must be sorted according to $r_i$), and the equations of the motion assume the form:
\begin{equation}
\begin{cases}
 \frac{\displaystyle \mathrm{d} r_i} {\displaystyle \mathrm{d}t}= \frac{\displaystyle p_{r,i}}{\displaystyle m_i},  \\
  \frac{\displaystyle \mathrm{d} p_{r,i}}{\displaystyle \mathrm{d}t} = q_i{E}_{r,i}(r_1,r_2,...,r_N)+\frac{\displaystyle p^2_{\varphi,i}}{\displaystyle m_i r_i^3}\label{pp2Dr},
\end{cases}
\end{equation}
in which the $p_{\varphi,i}$'s are constants of the motion and they are fixed by the initial conditions. This last formulation is the most convenient in terms of memory usage and computational effort. However, the presence of the term $p^2_{\varphi}/(m r^3)$ in Eqs. (\ref{pp2Dr}) require a special care when $r \rightarrow 0$. All things considered, the second formulation represents a good compromise in terms of computational efficiency and simplicity.

\subsection{Interaction between shells}
Due to symmetry, each computational particle can be regarded as a spherical surface (a ``shell") on which the electric charge is distributed uniformly.
The points on the surface move according to different trajectories, all sharing the same radial coordinate, $r(t)$, and the same angular momentum $p_{\varphi}$.
For simplicity, a system made of only two shells (having charge $q_1$ and $q_2$ and radii $r_1$ and $r_2$, with $r_1<r_2$) is considered now. As the electric field is given by
\begin{equation}
E(r)=\left\{\begin{array}{cccccc}
0,& r<r_{1},\\
\dfrac{q_1}{r^2},& r_1<r<r_{2},\\
\dfrac{q_1+q_2}{r^2},& r>r_{2},
\end{array}\right.
\label{pir}
\end{equation}
the electrostatic energy $U$ can be readily evaluated, as
\begin{equation}
    U(r_1,r_2)= \int_{\mathbb{R}^3} \frac{E^2}{8 \pi} \, \operatorname{d^3}\mathbf{x}= \frac{q_1^2}{2r_1}+\frac{q_2^2+2q_1q_2}{2r_2}.
\end{equation}
If $r_1$ is changed of $\delta r_1$, the change $-\delta U$ of the energy is equal to the work $qE_1 \cdot \delta r_1$ of the field on the shell itself. In other terms, one has:
\begin{equation}
    {E}_{1}=-\frac{1}{q_1} \frac{\partial U}{\partial r_1}= \frac{\frac{1}{2}q_1}{r_1^2}. \label{ee1}
\end{equation}
Similarly, the field acting on the second shell can be calculated as
\begin{equation}
    {E}_{2}=-\frac{1}{q_2} \frac{\partial U}{\partial r_2}= \frac{q_1+\frac{1}{2}q_2}{r_2^2}. \label{ee1}
\end{equation}
In both cases, the value of the electric field is in agreement with the rule ``${\sum\limits_{j=1}^{i-1}  q_j+\tfrac{1}{2}q_i}$", which was introduced
previously.\\ Now the dynamics of the two shells is considered. If there is no crossing (i.e., no collisions) between shells,  $r_1$ is always smaller than $r_2$ and one has
\begin{equation}
\begin{array}{cccccc}
\dfrac{\mathrm{d} p_1} {\mathrm{d}t}= q_1\dfrac{\frac{1}{2}q_1}{r_1^2}&, &\dfrac{\mathrm{d} p_2} {\mathrm{d}t}= q_2\dfrac{q_1+\frac{1}{2}q_2}{r_2^2}.
\end{array}\label{fus}
\end{equation}
Here only radial motion is considered for simplicity (i.e., $p_{\varphi}=0$ for both shells). The two equations (\ref{fus}) can be also written as
\begin{equation}
\begin{cases}
 \dfrac{\mathrm{d} p_1} {\mathrm{d}t}= -\dfrac{\partial}{\partial r_1}\left(\dfrac{\frac{1}{2}q_1^2}{r_1}\right),  \\
  \dfrac{\mathrm{d} p_2}{\mathrm{d}t} = -\dfrac{\partial}{\partial r_2}\left(\dfrac{q_1q_2+\frac{1}{2}q_2^2}{r_2}\right) \label{partial2Dr},
\end{cases}
\end{equation}
from which one immediately obtains
\begin{equation}
\begin{cases}
 \dfrac{p_1^2} {2m_1}+\dfrac{\frac{1}{2}q_1^2}{r_1}=\text{Const},  \\
 \dfrac{p_2^2} {2m_2}+\dfrac{q_1q_2+\frac{1}{2}q_2^2}{r_2}=\text{Const}.
\label{const2Dr}
\end{cases}
\end{equation}
As the two shells continue to expand, the asymptotic kinetic energy for $t \rightarrow +\infty$, $\mathcal{E}(+\infty)$, of the two shells can be readily evaluated, as
\begin{equation}
\begin{cases}
 \mathcal{E}_1(+\infty)=\mathcal{E}_1(0)+\dfrac{\frac{1}{2}q_1^2}{r_1(0)},\\
\mathcal{E}_2(+\infty)=\mathcal{E}_2(0)+\dfrac{q_1q_2+\frac{1}{2}q_2^2}{r_2(0)}.
\label{ener2Dr}
\end{cases}
\end{equation}
Now, the case of collision is considered. When $t=t_c$ one has $r_1(t_c)=r_2(t_c)=r_c$, and for $t>t_c$ the shell $\#$1 overtakes the shell $\#$2. Therefore, Eqs. (\ref{fus}-\ref{const2Dr}) are valid only for $t<t_c$. For $t>t_c$, Eqs. (\ref{fus}) must be replaced by
\begin{equation}
\begin{cases}
\dfrac{\mathrm{d} p_1} {\mathrm{d}t}= q_1\dfrac{q_2+\frac{1}{2}q_1}{r_1^2},  \\
 \dfrac{\mathrm{d} p_2} {\mathrm{d}t}= q_2\dfrac{\frac{1}{2}q_2}{r_2^2}
\label{fustwo}
\end{cases}
\end{equation}
(they are obtained by simply exchanging indices 1 and 2), from which one finally obtains
\begin{equation}
\begin{cases}
 \dfrac{p_1^2} {2m_1}+\dfrac{q_1q_2+\frac{1}{2}q_1^2}{r_1}=\text{Const},  \\
 \dfrac{p_2^2} {2m_2}+\dfrac{\frac{1}{2}q_2^2}{r_2}=\text{Const}.
\label{const2Drmod}
\end{cases}
\end{equation}
In the case of collision, in order to evaluate the new asymptotic energy, $\mathcal{E}^{\prime}(+\infty)$, both Eqs. \ref{const2Dr} (for $t<t_c$) and Eqs. \ref{const2Drmod} must be considered:
\begin{equation}
\begin{cases}
 \mathcal{E}^{\prime}_1(t_c)=\mathcal{E}_1(0)+\dfrac{\frac{1}{2}q_1^2}{r_1}-\dfrac{\frac{1}{2}q_1^2}{r_c}= \mathcal{E}_1(+\infty)-\dfrac{\frac{1}{2}q_1^2}{r_c},\\\\
\mathcal{E}^{\prime}_2(t_c)=\mathcal{E}_2(0)+\dfrac{q_1q_2+\frac{1}{2}q_2^2}{r_2}-\dfrac{q_1q_2+\frac{1}{2}q_2^2}{r_c}= \mathcal{E}_2(+\infty)-\dfrac{q_1q_2+\frac{1}{2}q_2^2}{r_c}
\label{enerv0},
\end{cases}
\end{equation}
and
\begin{equation}
\begin{cases}
 \mathcal{E}^{\prime}_1(+\infty)=\mathcal{E}^{\prime}_1(t_c)+\dfrac{q_1q_2+\frac{1}{2}q_1^2}{r_c}=\mathcal{E}_1(+\infty)+\dfrac{q_1q_2}{r_c},\\
\mathcal{E}^{\prime}_2(+\infty)=\mathcal{E}^{\prime}_2(t_c)+\dfrac{\frac{1}{2}q_2^2}{r_c}= \mathcal{E}_2(+\infty)-\dfrac{q_1q_2}{r_c}
\label{enerv1}.
\end{cases}
\end{equation}
In other terms, the collision produces an increase $\Delta\mathcal{E}=q_1q_2/r_c$ in the energy of the shell $\#$1, and a corresponding decrease $-\Delta\mathcal{E}$ for the shell $\#$2.
In a typical plasma expansion, the energy $\mathcal{E}$ of a shell is of the order of $qQ/R$, being $Q$ the total charge and $R$ the initial plasma radius.
Being $\Delta\mathcal{E} \sim q^2/R$ for a single collision, one can conclude that the ``plasma parameter" $\Delta\mathcal{E}/\mathcal{E}$ for a set on $N$
shells will be of the order of $q/Q=1/N$. In practice, for typical values of the number of computational particles, the system can always be regarded as collisionless.

\begin{figure}[!h]
\centering
\includegraphics[width=0.65\linewidth]{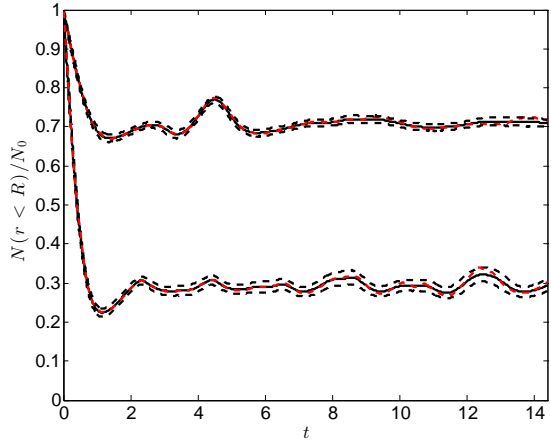}
\caption{Time evolution of the fraction of electrons inside the ion sphere for two different normalized temperature, $T=0.0431, 0.431$. For each value of $T$, ensemble averages (full black line) and standard deviation ranges (dashed black lines) are reported for $N=10^3$ shells and 300 simulations with different initial conditions, together with reference results provided by a simulation with $N=10^6$ shells (dashed red line).}
\label{s1}
\end{figure}

\begin{figure}[!h]
\centering
\includegraphics[width=0.65\linewidth]{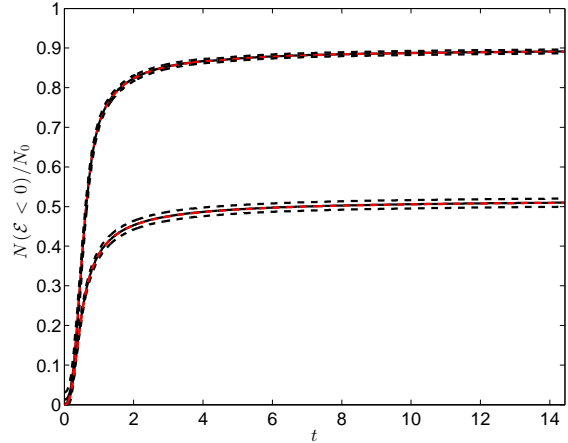}
\caption{Time evolution of the fraction of trapped electrons for the same case of Fig. \ref{s1}.}
\label{s2}
\end{figure}

\subsection{Results}
Some typical results are reported in the following. In all the calculations, suitable normalization for the physical quantities has been used such that the total charge, the total mass of the plasma and the initial radius $R$ are all equal to 1. Three cases are
considered: 1) the electron expansion in a spherical plasma
\cite{Peano-PRL-2006}; 2) the expansion of a plasma made of a
mixture of two ion species \cite{Boella-JPP-2016}; 3) the formation
of shocks in Coulomb explosions \cite{peanosilva}. Figures
\ref{s1} and \ref{s2} refer to the early stage of the electron
expansion in a spherical plasma. It is assumed that electrons and
positive ions are initially distributed uniformly in a sphere of
radius $R$. Initially, electrons have Maxwellian velocity
distribution with temperature $T$ and positive ions are considered
at rest during all the transient. Calculations have been performed
both with a reduced ($N\simeq10^3$) and with a high number of shells
($N\simeq10^6$), in order to obtain reference results. The initial
phase-space distribution of the electrons was generated by using
random numbers, so for a small number of particles the results will
depend on the particular choice of positions and velocities. For
this reason, the same calculation has been repeated for 300 times
(with different initial conditions, all corresponding to the same
physical situation) in order to obtain the mean behavior and the
distribution of the physical quantities (as performed in
\cite{Dangola-PoP-2014}). In Figs. \ref{s1} and \ref{s2}, the time
evolution of the number of electrons inside the ion sphere (i.e.,
with $r \le R$) and of the fraction of trapped electrons (i.e., with
total energy $\tfrac{p^2}{2m}-e\Phi(r) \le 0$) are reported, respectively.
As can be observed, the shell method provides excellent results,
even with a reduced set of particles.
\begin{figure}[!h]
\centering
\includegraphics[width=0.65\linewidth]{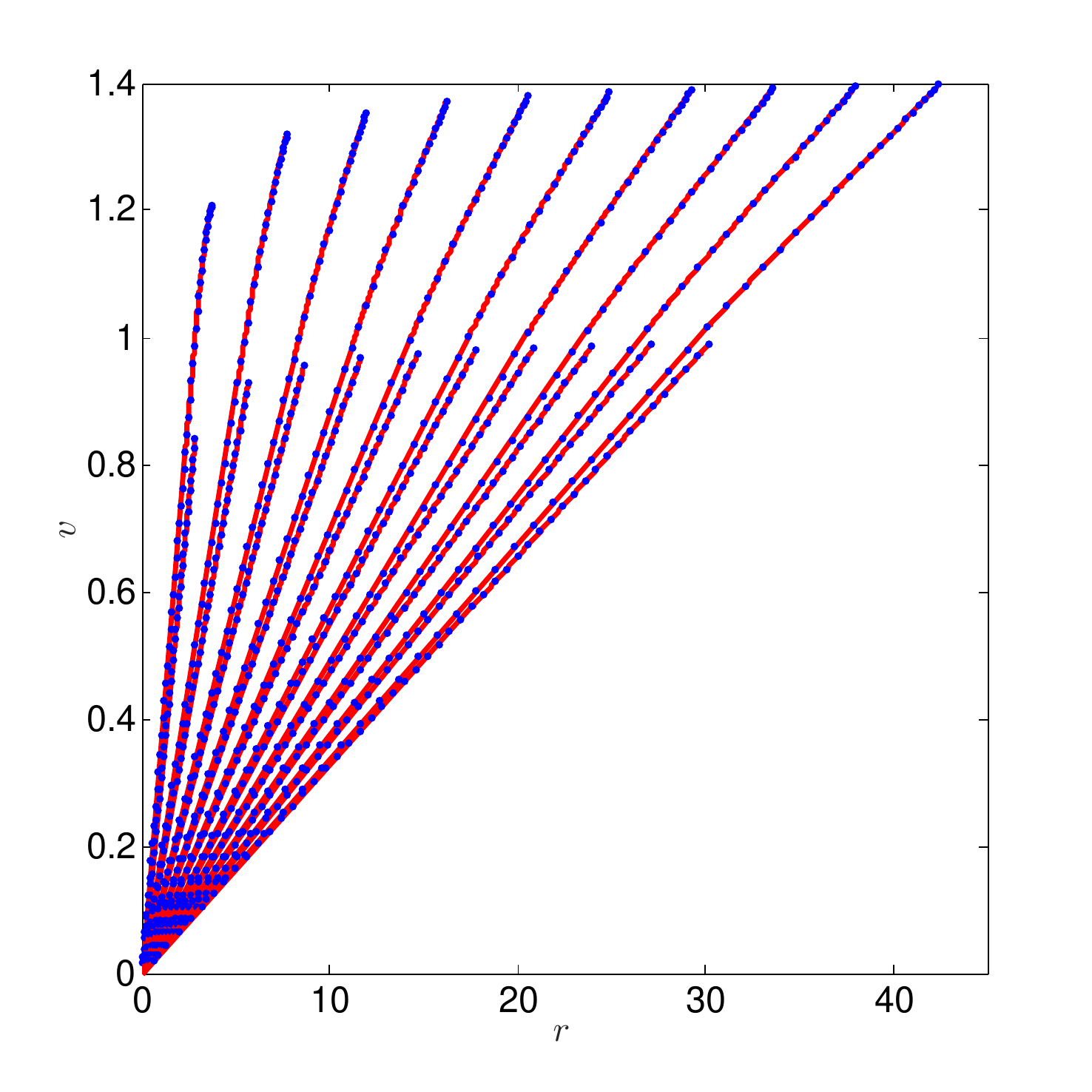}
\caption{Phase-space distributions of a mixture with $m_1/m_2=2/3$ and $q_1=q_2$ at different times ($t=3\div 31$). Results obtained with the shell method (blue dots) are compared with the analytic solution (red solid lines).}
\label{fig:DT_gusci1}
\end{figure}

\begin{figure}[!h]
\centering
\includegraphics[width=0.65\linewidth]{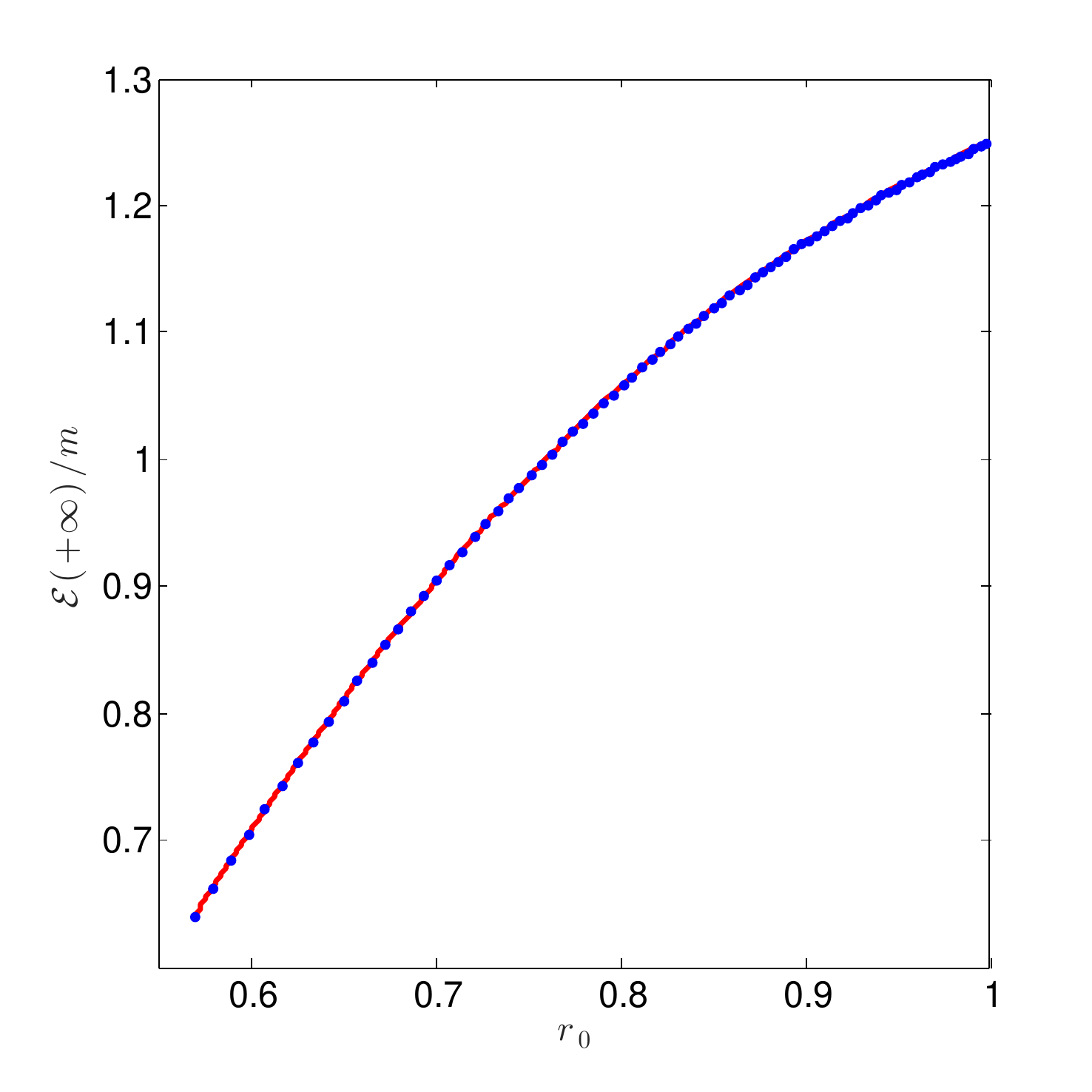}
\caption{$\mathcal{E}(t \rightarrow +\infty)/m$ of the light ions as a function of their initial radial coordinate, $r_0$, for the case of Fig. \ref{fig:DT_gusci1}. Results obtained with the shell method  (blue dots) are compared with the analytic solution (red line).}
\label{fig:DT_gusci2}
\end{figure}

\begin{figure}[!h]
\centering
\includegraphics[width=0.65\linewidth]{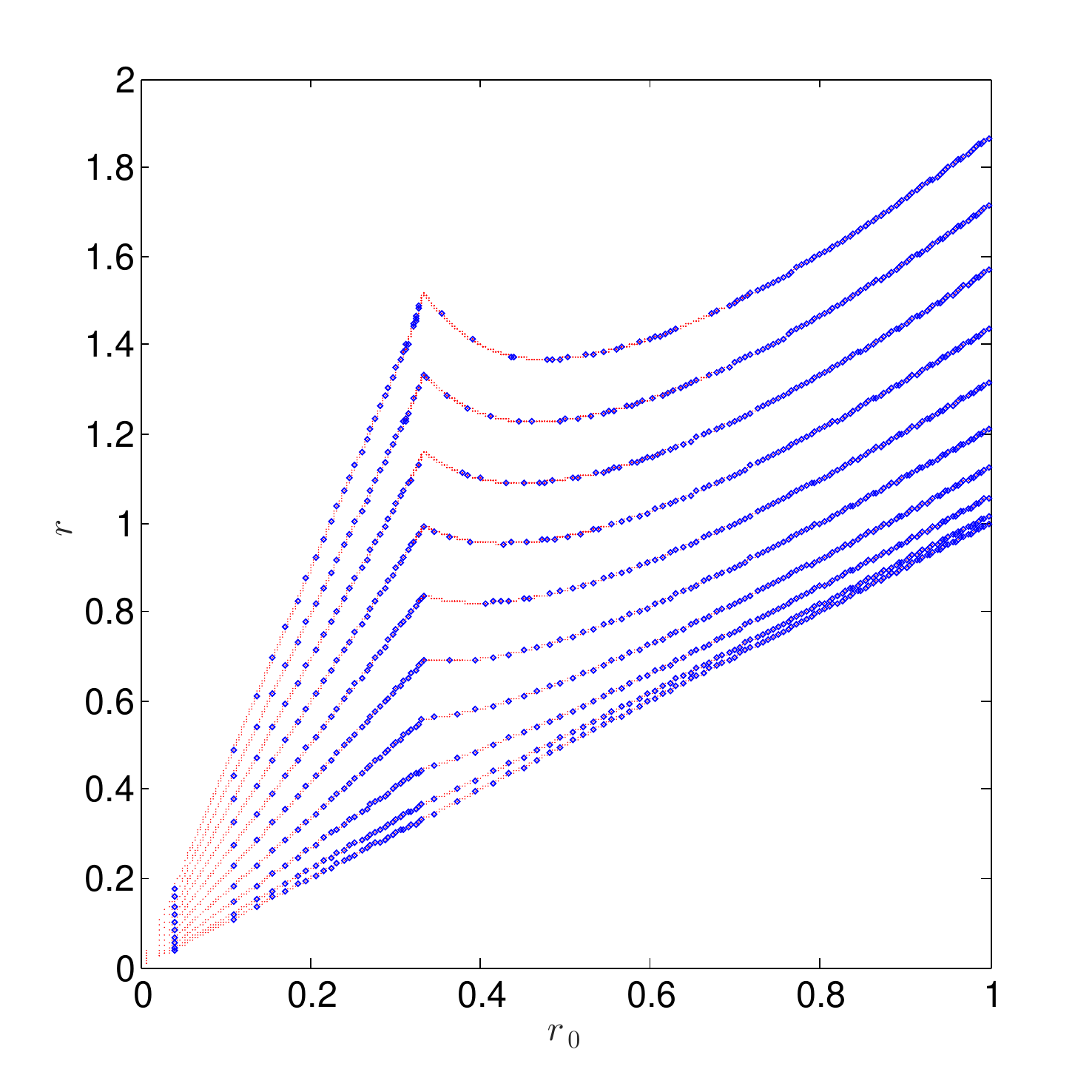}
\caption{Radial coordinate, $r$, at different times ($t=0 \div 1.47$) as a function of their initial position, $r_0$, for a single-species ion plasma with a non uniform initial density distribution. In the simulations, $n(r,0)=n_1$ when $r<R/3$ and $n_2$ when $r \in \left[ \tfrac{R}{3}, R\right]$, with $n_1/n_2=8$. Results for $10^4$ shells (blue dots) are compared with those obtained with $10^6$ shells (red line). 
}
\label{fig:gusci3_1}
\end{figure}

\begin{figure}[!h]
\centering
\includegraphics[width=0.65\linewidth]{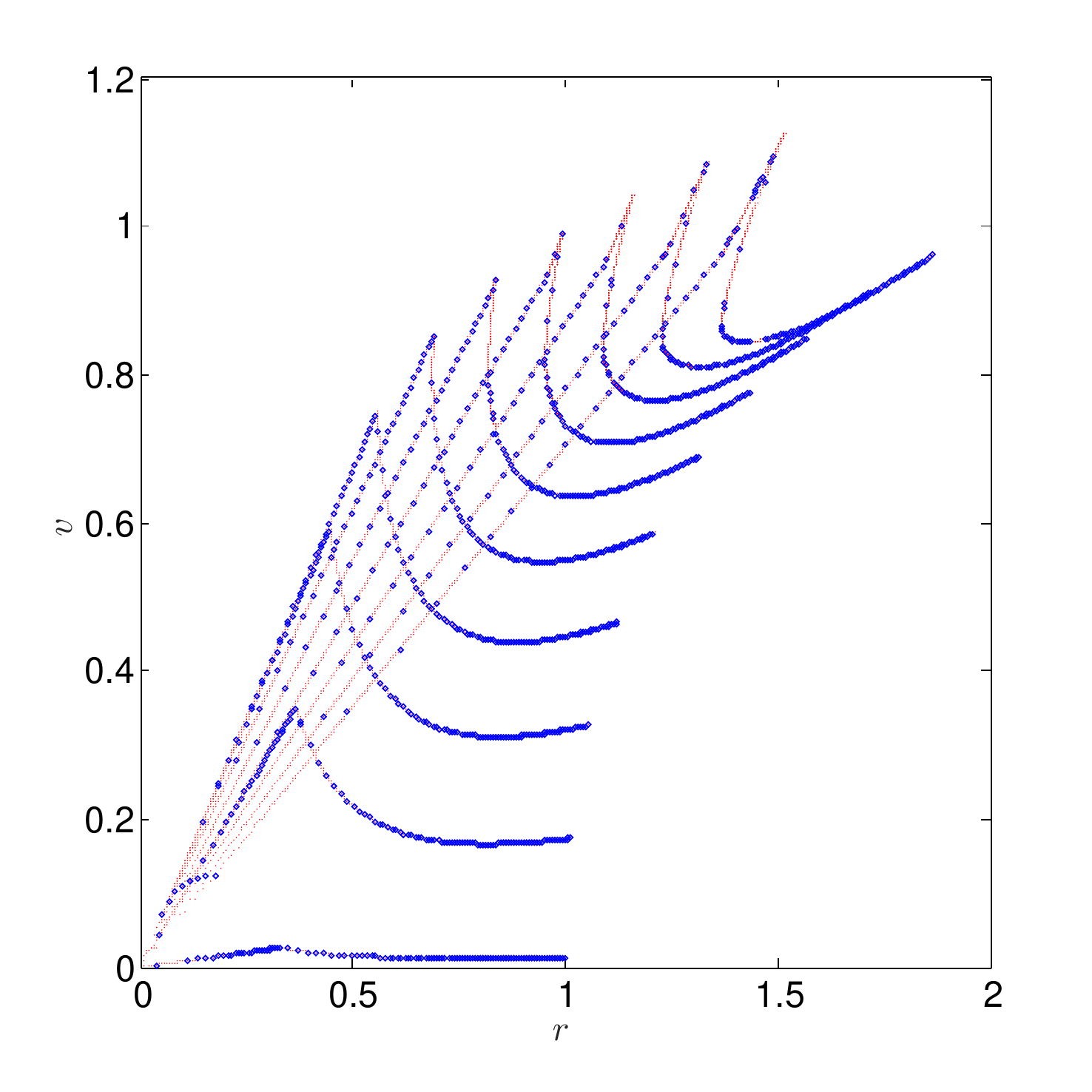}
\caption{Ion phase-space distribution at different times ($t=0  \div 1.47$) for the same case of Fig. \ref{fig:gusci3_1}. Results for $10^4$ shells (blue dots) are compared with those obtained with $10^6$ shells (red line).}
\label{fig:gusci3_2}
\end{figure}
The second set of results (Figs. \ref{fig:DT_gusci1} and \ref{fig:DT_gusci2}) refers to the acceleration of an ion plasma
made of a mixture of two different species. In this case, analytic solutions for the problem exist \cite{Boella-JPP-2016}
and can be used as a reference. The two species ($m_1/m_2=2/3, q_1=q_2$) are initially at rest and the ions are accelerated
by electrostatic repulsion. In Fig. \ref{fig:DT_gusci1} the phase-space distribution for the two species, calculated with
the shell method and using $10^3$ computational particles, is reported at different times and compared with analytic results.
Figure \ref{fig:DT_gusci2} shows  $\mathcal{E}(t \rightarrow +\infty)/m$ of the light ions as a function of their initial
radial coordinate, $r=r_0$. This curve is important in order to determine the asymptotic energy spectrum,
$\tfrac{\operatorname{d} N}{\operatorname{d} \mathcal{E}}$, of the ions (considering
that $\Delta \mathcal{E} = \tfrac{\operatorname{d} \mathcal{E}}{\operatorname{d} r_0} \Delta r_0$ and $\Delta N=4\pi r_0^2 n_0 \Delta r_0$). The two figures show the excellent agreement between numerical and analytic results. \\
The third case here considered concerns the shock formation in a
Coulomb explosion \cite{Peano-PRL-2006, Peano-PRE-2007}. The
phenomenon arises when the initial ion distribution is not uniform,
in particular if the inner density is larger respect to the outer
one. In fact, in this case the electric field has a maximum inside
the plasma region (while it depends linearly on $r$ if the ion
density is constant) and consequently inner particles acquire higher
kinetic energy with respect to the outer ones and can ``overtake"
them. In the situation considered in Figs. \ref{fig:gusci3_1} and
\ref{fig:gusci3_2}, an ion plasma made of only one species presents
two regions with different density for $t=0$. Figure
\ref{fig:gusci3_1} reports the value of the radial coordinate
$r(r_0,t)$ of the ions as a function of their initial radius, $r_0$,
for different times, while in Fig. \ref{fig:gusci3_2} the
phase-space distribution is plotted. The results here reported show
the ability of the shell method to analyze cases in which the
density, in theory, may become infinite in some point; in fact,
results obtained with a relative low ($10^4$) and with a very large
($10^6$) number of shells are in perfect agreement.

\section{The ring method} \label{ring}

\begin{figure}
    \centering
    \includegraphics[width=0.7\linewidth]{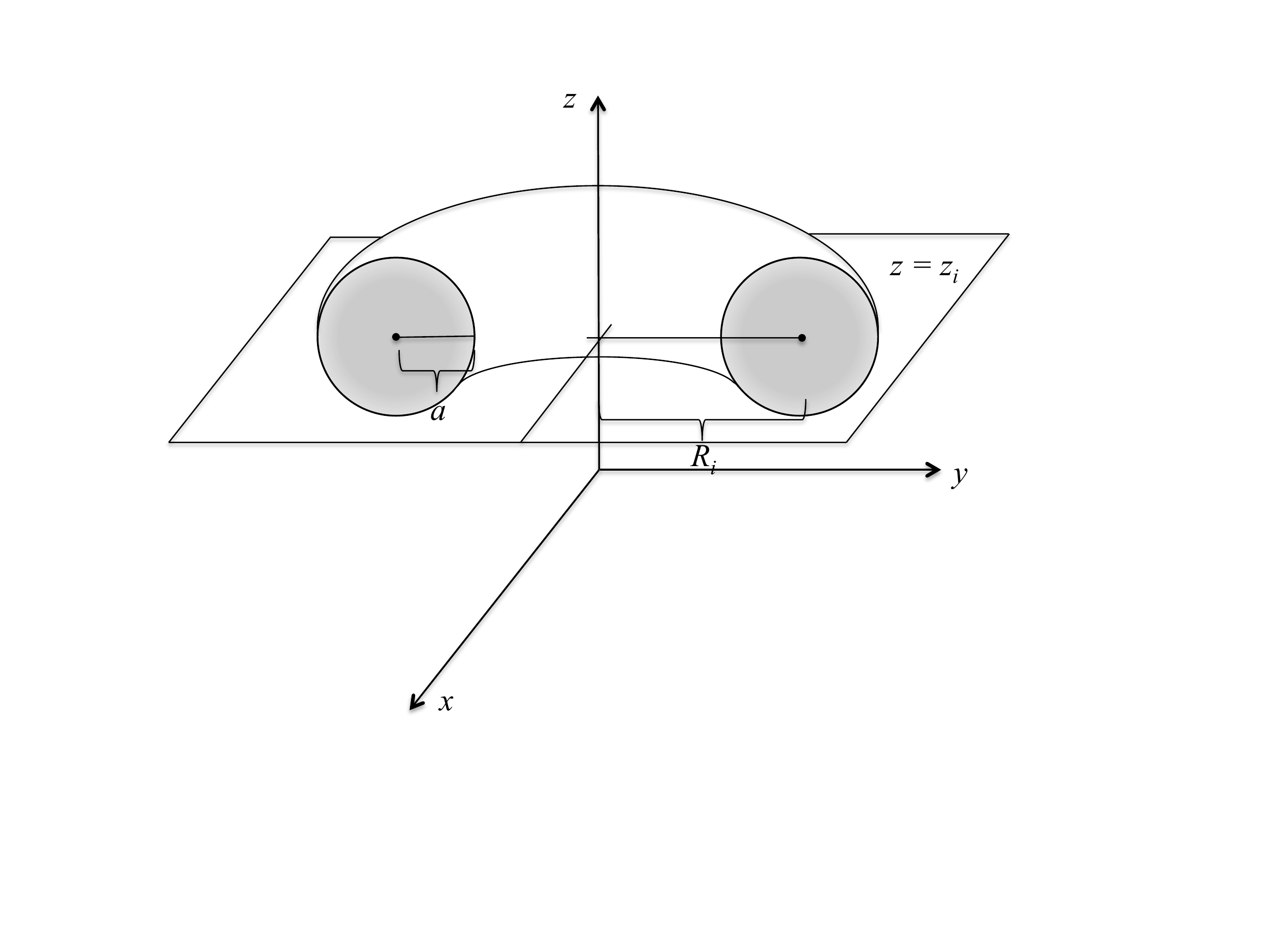}
      \caption{Scheme of a torus.}
    \label{fig:app1}
\end{figure}

In the case of axial symmetry the fundamental ``brick" for a $N$-body technique is a ring. More precisely, tori having circular
cross section (of radius $a$) are considered here. The tori shares the same axis of symmetry (the $z$ axis) and are characterized by their radii, $R_i$, and axial coordinates, $z_i$ (as in Fig. \ref{fig:app1}).
When $N$ tori are considered, the electrostatic energy of the system can be written as:
\begin{equation}
U = \tfrac{\displaystyle 1}{\displaystyle 2}\sum_{i \ne j} q_i q_j \varphi_{ring}(R_i,R_j,z_i-z_j)+\sum_{i =1}^N q_i^2 U_{torus}(R_i,a),
\label{U_tt}
\end{equation}
where $\varphi_{ring}(R,R^{\prime},z^{\prime})$ is the potential generated by a unit charge distributed on a ring (i.e., a torus with $a=0$) of
radius $R$ laying on the $xy$ plane in a point of polar coordinates $(R^{\prime},z^{\prime})$, while $U_{torus}(R,a)$ is the potential energy of a torus of unitary charge.
The potential $\varphi_{ring}(R;R^{\prime},z^{\prime})$ can be evaluated\footnote{As a generic point of the ring has coordinates $(R\cos(\vartheta),R\sin(\vartheta),0)$
and the point where the potential has to be evaluated has coordinates $(R^{\prime},0,z^{\prime})$, the potential $\varphi_{ring}$ can be written as
\begin{equation}
\varphi_{ring}= \frac{1}{2\pi} \int_{0}^{2\pi}  \frac{1}{(R^2+R^{\prime 2}+z^{\prime 2}-2RR^{\prime}\cos \theta)^{1/2}}\operatorname{d}\theta,
\label{Phi_anelliring}
\end{equation}
By introducing the new integration variable $\alpha=\frac{\theta}{2}-\frac{\pi}{2} $, the formula for $\varphi_{ring}$ becomes:
\begin{equation}
\varphi_{ring}=\frac{2 q}{\pi}\int_{0}^{\pi/2} \frac{1}{\left[(R+R^{\prime})^2+z^{\prime 2}-4RR^{\prime}\sin^2\alpha\right]^{1/2}}\operatorname{d}\alpha,
\end{equation}
from which Eq. (\ref{Phi_PO_app}) immediately follows.
} in terms of the complete elliptic integral of the first kind \cite{abra}:
\begin{equation}
K[x]=\int_{0}^{\pi/2} \frac{\operatorname{d}\alpha}{(1-x\sin^2\alpha)^{1/2}},
\end{equation}
as
\begin{equation}
\varphi_{ring}(R;R^{\prime},z^{\prime})=\frac{2K[\xi]}{\pi s},
\label{Phi_PO_app}
\end{equation}
being
\begin{equation}
s=[(R+R^{\prime})^2+ z^{\prime 2}]^{1/2},\;\; \text{    } \;\; \xi=\frac{4RR^{\prime}}{s^2}.
\label{xi_s}
\end{equation}

\begin{figure}
    \centering
    \includegraphics[width=0.65\linewidth]{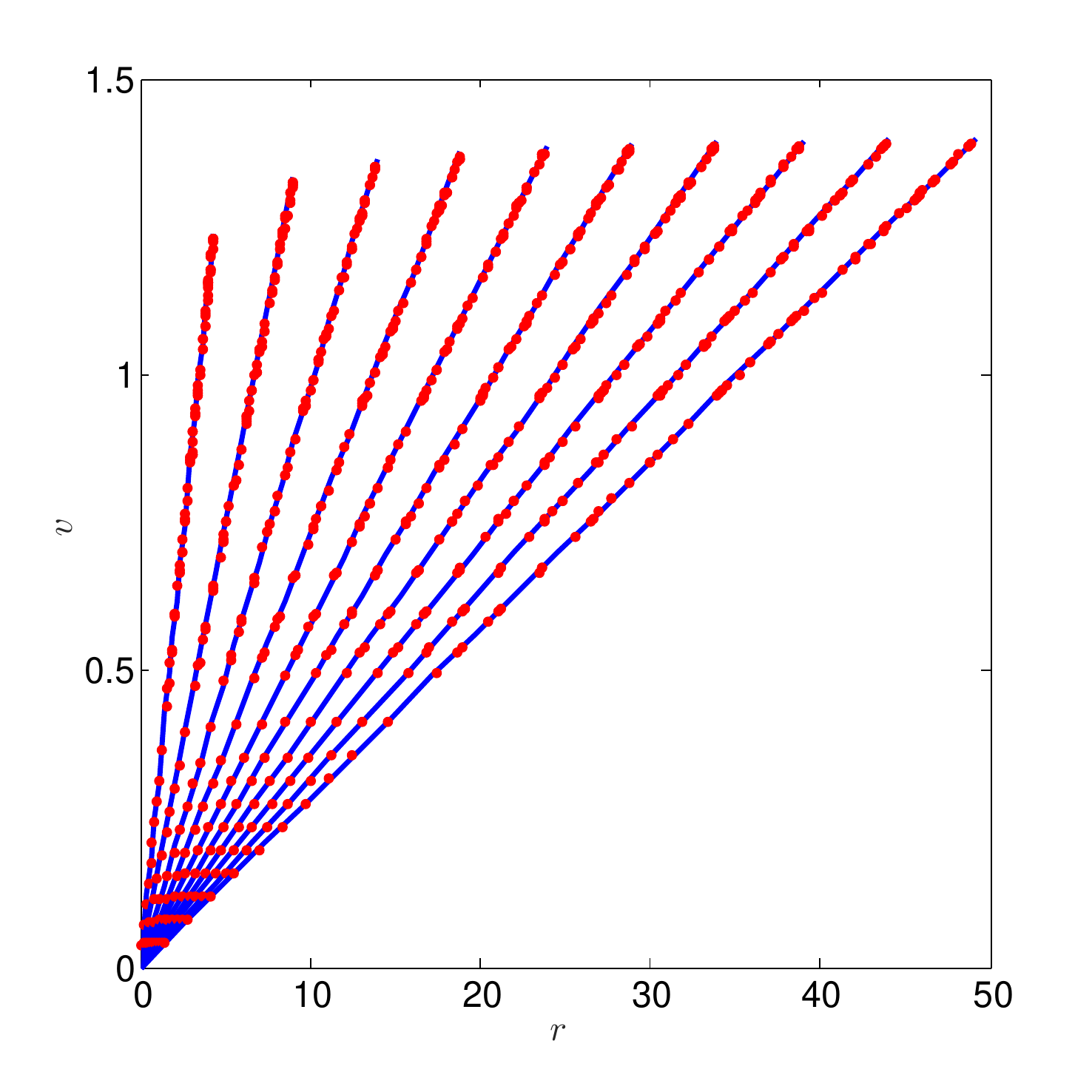}
       \caption{Phase-space distribution at different times ($t=4 \div 36$) of a spherical ion plasma in the case of ring loading with method 1.  Results obtained with the ring method (blue dots) are compared with the analytic solution (red lines).}
    \label{fig:sfera_random_1}
\end{figure}

\begin{figure}
    \centering
    \includegraphics[width=0.65\linewidth]{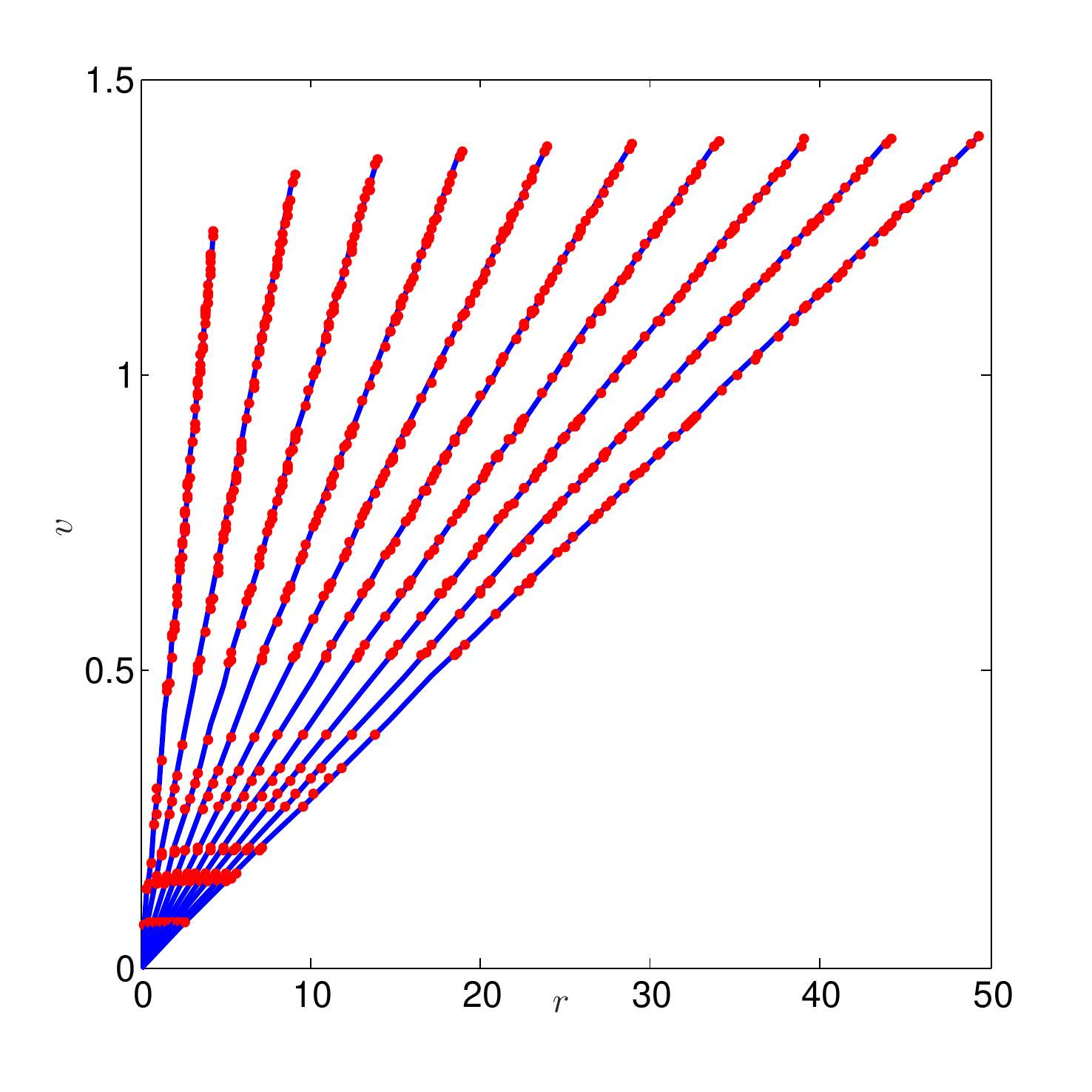}
    \caption{Same as Fig. \ref{fig:sfera_random_1} in the case of ring loading with method 2 ($t=4 \div 36$).  Results obtained with the ring method (blue dots) are compared with the analytic solutions (red lines).}
    \label{fig:sfera_spaced_1}
\end{figure}

\begin{figure}
    \centering
    \includegraphics[width=0.65\linewidth]{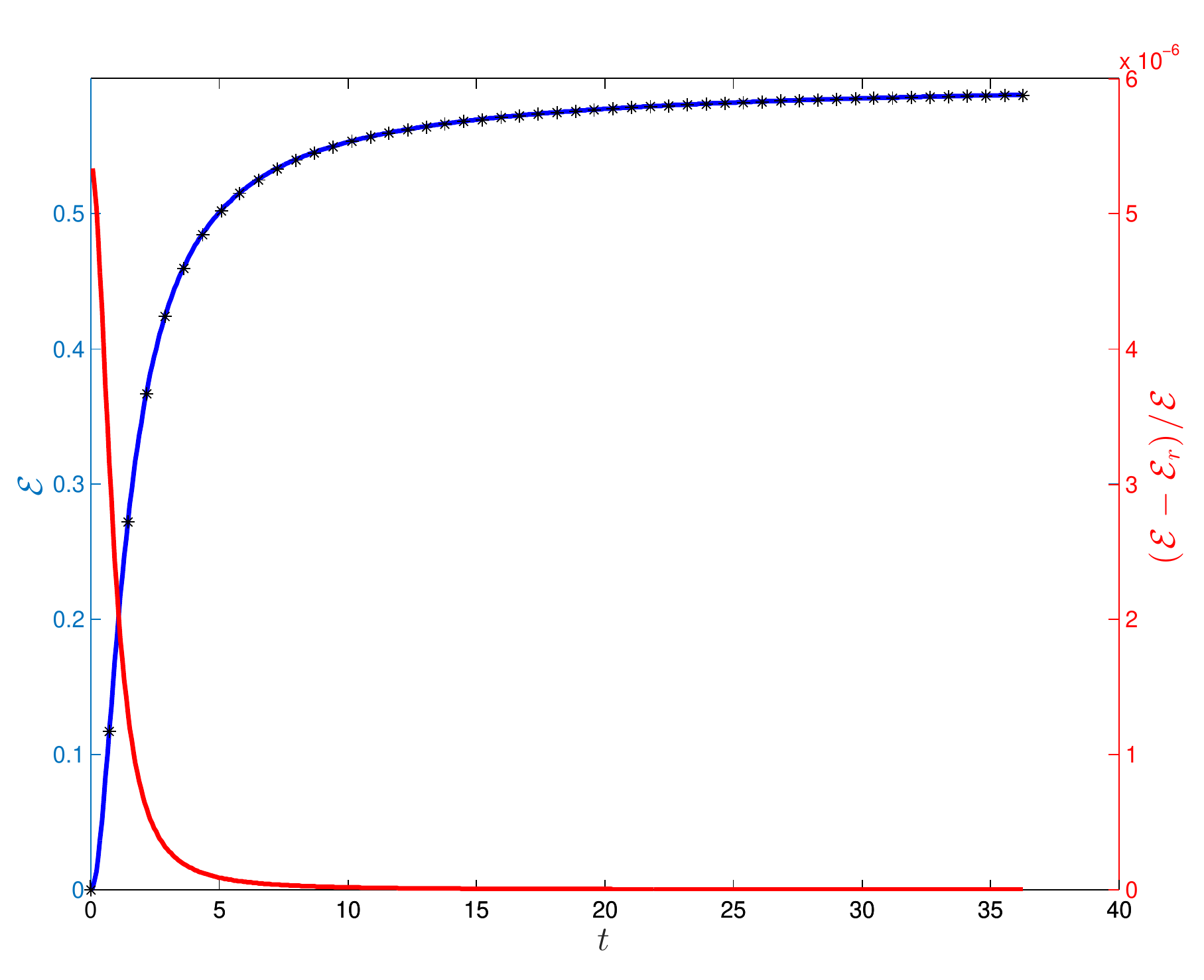}
       \caption{Time evolution of the total kinetic energy of the ions (blue line) and of the fraction of the perpendicular kinetic energy (red line) obtained with the ring method for the same case of Fig. \ref{fig:sfera_random_1} (method 1 for ring loading). Results obtained with the ring method are compared with the analytic solutions (black stars).}
    \label{fig:sfera_random_2}
\end{figure}

\begin{figure}
    \centering
    \includegraphics[width=0.65\linewidth]{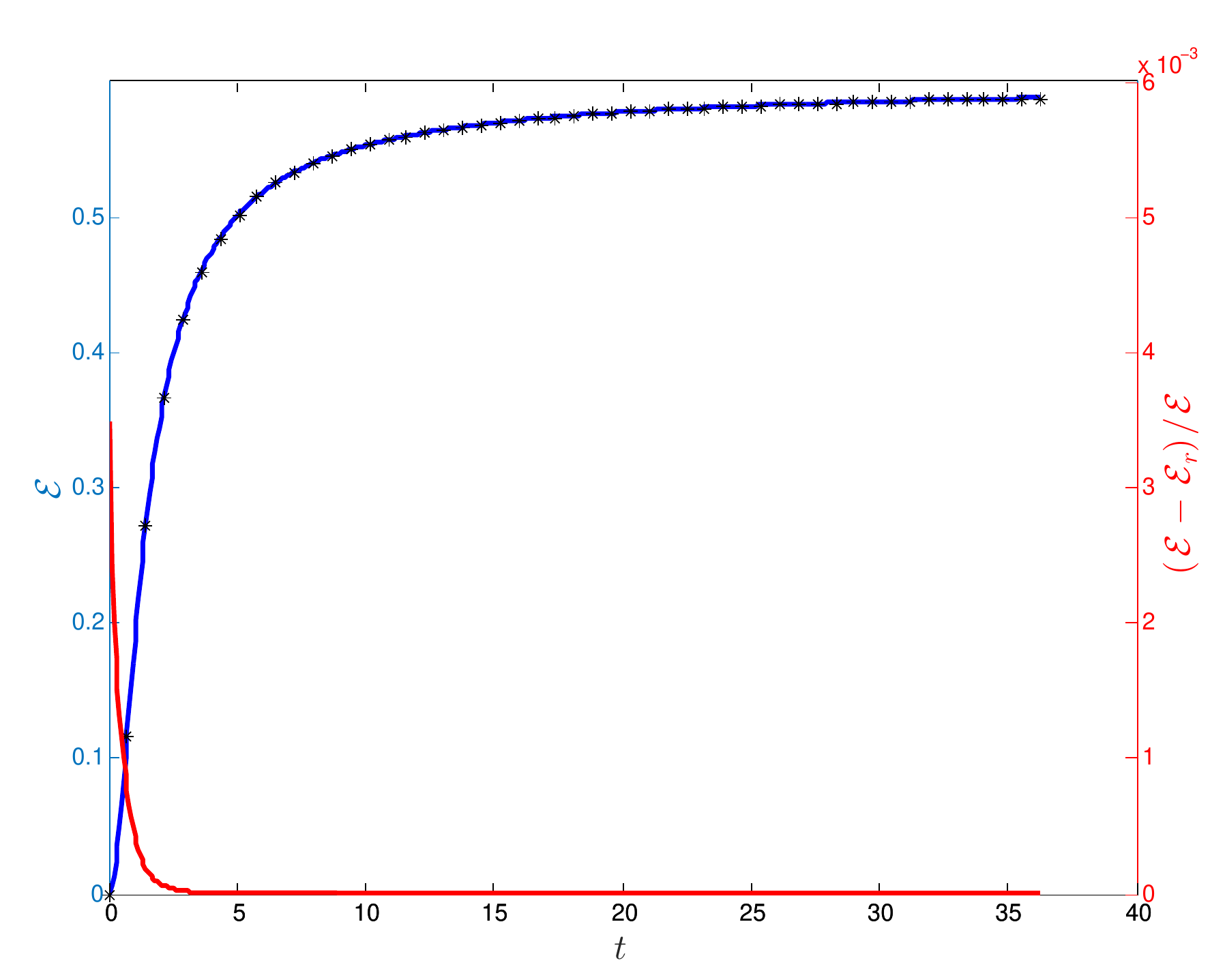}
    \caption{Same as Fig. \ref{fig:sfera_random_2}, using method 2 for ring loading.}
    \label{fig:sfera_spaced_2}
\end{figure}

The calculation of $U_{torus}(R,a)$ is reported in detail in the Appendix. For the case of interest in which $a \ll R$, one has:
\begin{equation}
U_{torus}(R,a) \widesim[2]{} -\frac{1}{2 \pi R}\left[\log\left(\frac{a}{ 8 R}\right)-\frac{1}{4}\right].
\label{Utoro}
\end{equation}
From Eq. (\ref{Utoro}), it can be noticed that $U_{torus}$ diverges for $a\rightarrow 0$, and this is the reason why tori are considered and
not simply rings. Instead, in calculating the interaction energy between tori, the value of $\varphi_{ring}$ is employed, as it is supposed that when $a \ll R$ the energy of two tori or two rings is essentially the same.\\
Now, the equations of the motion for the set of rings are derived. In order to write the Lagrangian of the system, the kinetic energy
\begin{equation}
\sum_{i=1}^N\frac{m_i}{2}\left(\dot{R}_i^{2}+\dot{z}_i^{2}+R_i^{2}\dot{\varphi}_i^{2}\right)
\label{lagring}
\end{equation}
must be considered. By introducing the momenta $p_{R,i}$, $p_{z,i},$ and $p_{\varphi,i}$:
\begin{equation}
p_{R,i}=m_i \dot{R}_i, \hspace{2cm} p_{z,i}=m_i \dot{z}_i, \hspace{2cm} p_{\varphi,i}=m_i R_i^2\dot{\varphi}_i,
\end{equation}
one finally obtains the Hamiltonian $\mathscr{H}$ of the $N$ interacting rings as:
\begin{equation}
\mathscr{H}=\sum_{i=1}^N\frac{1}{2m_i}\left(p_{R,i}^{2}+p_{z,i}^{2}\\
+\frac{p^2_{\varphi,i}}{R_i^{2}}\right)+\tfrac{\displaystyle 1}{\displaystyle 2}\sum_{i \ne j} q_i q_j \varphi_{ring}(R_i,R_j,z_i-z_j)+\sum_{i =1}^N q_i^2 U_{torus}(R_i,a),
\label{hring}
\end{equation}
and the equations of the motion:
\begin{equation}
\begin{cases}
\dfrac{\operatorname{d} R_{\alpha}} {\operatorname{d}t}= \dfrac{p_{R,\alpha}}{m_{\alpha}}, \hspace{1cm}  \dfrac{\operatorname{d} z_{\alpha}} {\operatorname{d}t}= \dfrac{p_{z,\alpha}}{m_{\alpha}}, \\&  \\
 \dfrac{\operatorname{d} p_{R,\alpha}} {\operatorname{d}t}= \dfrac{p^2_{\varphi,\alpha}}{m_{\alpha}R_{\alpha}^3}-\sum\limits_{\beta \ne \alpha} q_{\alpha} q_{\beta} \dfrac{\partial}{\partial R_{\alpha}}\varphi_{ring}(R_{\alpha},R_{\beta},z_{\alpha}-z_{\beta})-q_{\alpha}^2  \dfrac{\partial}{\partial R_{\alpha}} U_{torus}(R_{\alpha};a), \\&  \\
 \dfrac{\operatorname{d} p_{z,\alpha}} {\operatorname{d}t}= -\sum\limits_{\beta \ne \alpha} q_{\alpha} q_{\beta} \dfrac{\partial}{\partial z_{\alpha}}\varphi_{ring}(R_{\alpha},R_{\beta},z_{\alpha}-z_{\beta}).
\end{cases}
\label{eq34}
\end{equation}
The angular momenta $p_{\varphi,\alpha}$ are constants of the motion. The partial derivatives of $\varphi_{ring}$ can be readily evaluated considering that:
\begin{equation}
\frac{\operatorname{d} K[x]} {\operatorname{d}x}=\frac{E[x]-(1-x)K[x]}{2x(1-x)},
\label{EKelli}
\end{equation}
being $E[x]=\int_0^{\pi/2}(1-x\sin^2 \alpha)^{1/2}\mathrm{d}\alpha$ the complete elliptic integral of the second kind \cite{abra}.
Equations (\ref{eq34}) have been deduced by considering only electrostatic interaction in non relativistic limit. In principle, the method can be readily extended to include relativistic particles and magnetic field (with axial symmetry).
To test its accuracy, the ring method has been employed to simulate the expansion of an ion sphere of uniform density, for which a simple analytic solution exists. The same normalizaion of the physical quantities of Sect. 2.5 is used here. The initial ring distribution $\{R_i,z_i\}$ has been generated in two different ways: 1) by dividing the initial $[R,z]$ domain (i.e., a half circle of radius $R_0$) into a number $N$ of small squares, each corresponding to the cross section of a ring; 2) by suitably taking a set of $\{R_i,z_i\}$ in a random way in order to obtain a uniform charge density. The radius $a_i$ of the section of each ring has been chosen as proportional to $R_i$, i.e., $a_i=k \cdot R_i$. The constant $k$ has been determined by requiring the potential energy of the set of the rings to be equal to the exact value of the energy of the sphere. Figures \ref{fig:sfera_random_1}, \ref{fig:sfera_spaced_1} and \ref{fig:sfera_random_2}, \ref{fig:sfera_spaced_2} refer to
method 1 and method 2, for ring loading, respectively. In Figs. \ref{fig:sfera_random_1} and \ref{fig:sfera_spaced_1} the time evolution of the phase-space
distribution, as obtained with the ring method, is shown and it is compared with its analytical behavior. Figures \ref{fig:sfera_random_2} and \ref{fig:sfera_spaced_2}
show the total kinetic energy of the ions, $\mathcal{E}=\sum_{i=1}^N \tfrac{m_i}{2}\mathbf{v}_i^2(t)$, as a function of $t$; moreover, the behavior
of $[\mathcal{E}^{(t)}-\mathcal{E}_r(t)]/\mathcal{E}$, where $\mathcal{E}_r(t)=\sum_{i=1}^N \tfrac{m_i}{2}\left[\mathbf{v}_i^2(t) \cdot \mathbf{e}_{r,i}(t)\right]$ is
the kinetic energy due to the motion in radial direction, is also presented. Obviously, in the exact solution $\mathcal{E}_r(t) \equiv \mathcal{E}(t)$, so a value
of $\bigl\lvert\tfrac{\mathcal{E}-\mathcal{E}_r(t)}{\mathcal{E}}\bigl\lvert\ll1$ is expected. All the numerical results presented in Figs. \ref{fig:sfera_random_1}, \ref{fig:sfera_spaced_1}, \ref{fig:sfera_random_2}, \ref{fig:sfera_spaced_2} are in excellent agreement with the theory.


\begin{figure}
    \centering
       \includegraphics[width=0.65\linewidth]{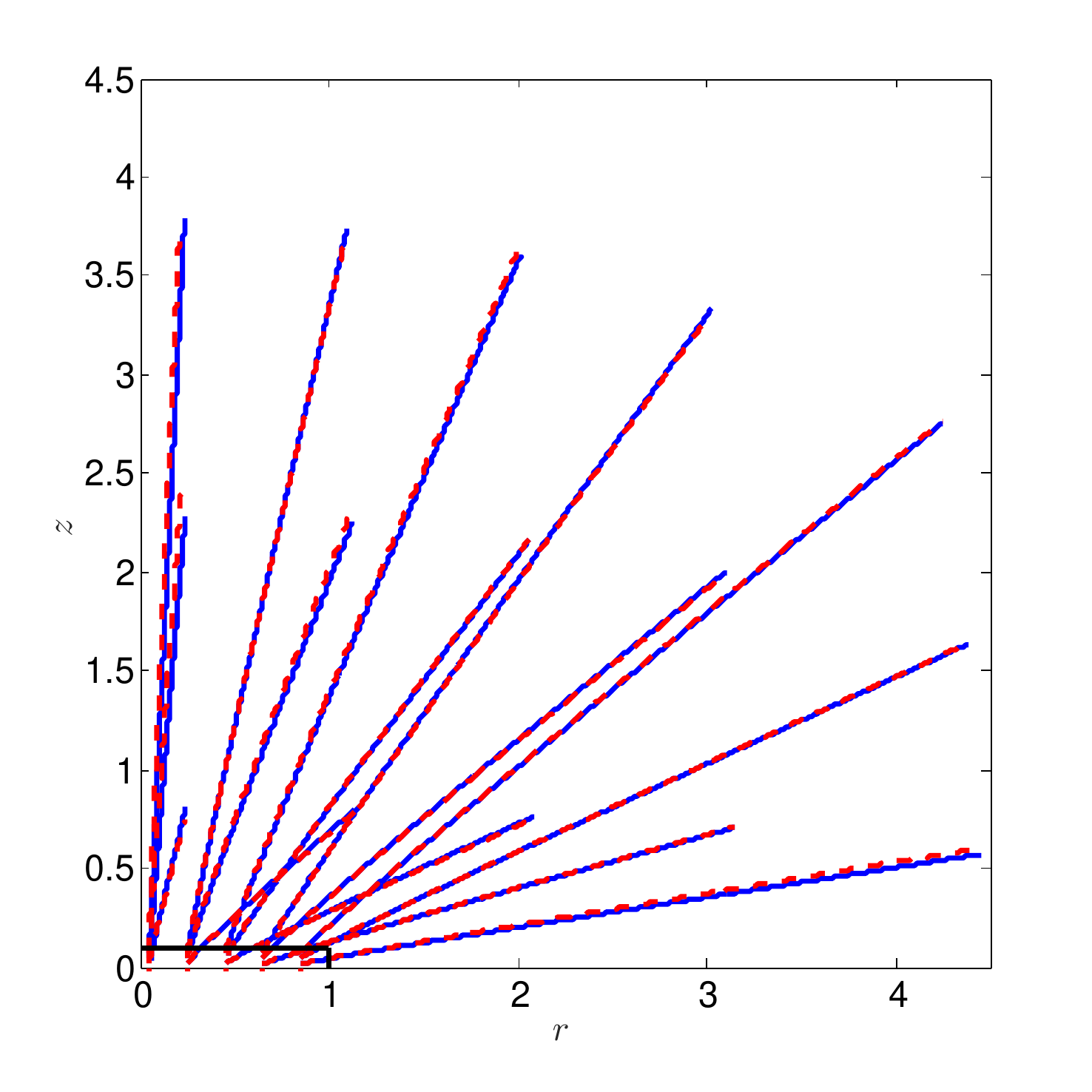}
    \caption{Particle trajectories for the Coulomb explosion of an ion plasma having initially a cylindrical shape (the ratio between initial radius $R$ and height $H$ is equal to 0.1) for $t=0 \div 4$. Results obtained with the ring method (blue lines) are compared with those obtained with the PIC method (red dotted lines).}
    \label{fig:sfera_random2_4}
\end{figure}
\begin{figure}
    \centering
    \includegraphics[width=0.65\linewidth]{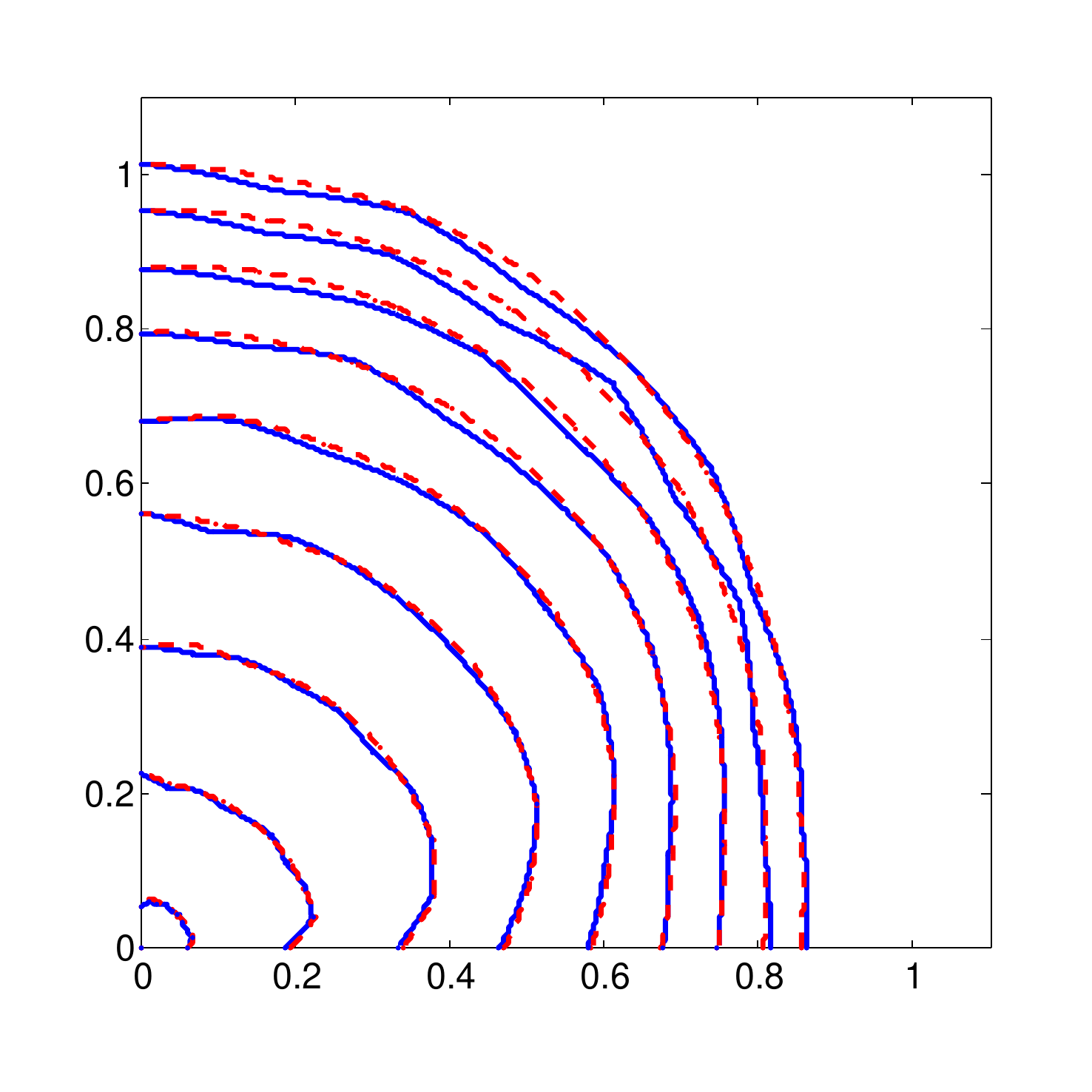}
    \caption{Angular distribution of $\mathcal{E}/m$ for the case of Fig. \ref{fig:sfera_random2_4} for $t=0 \div 4$. Results obtained with the ring method (blue lines) are compared with those obtained with the PIC method (red dotted lines).}
    \label{fig:sfera_random2_3}
\end{figure}

\begin{figure}
    \centering
       \includegraphics[width=0.65\linewidth]{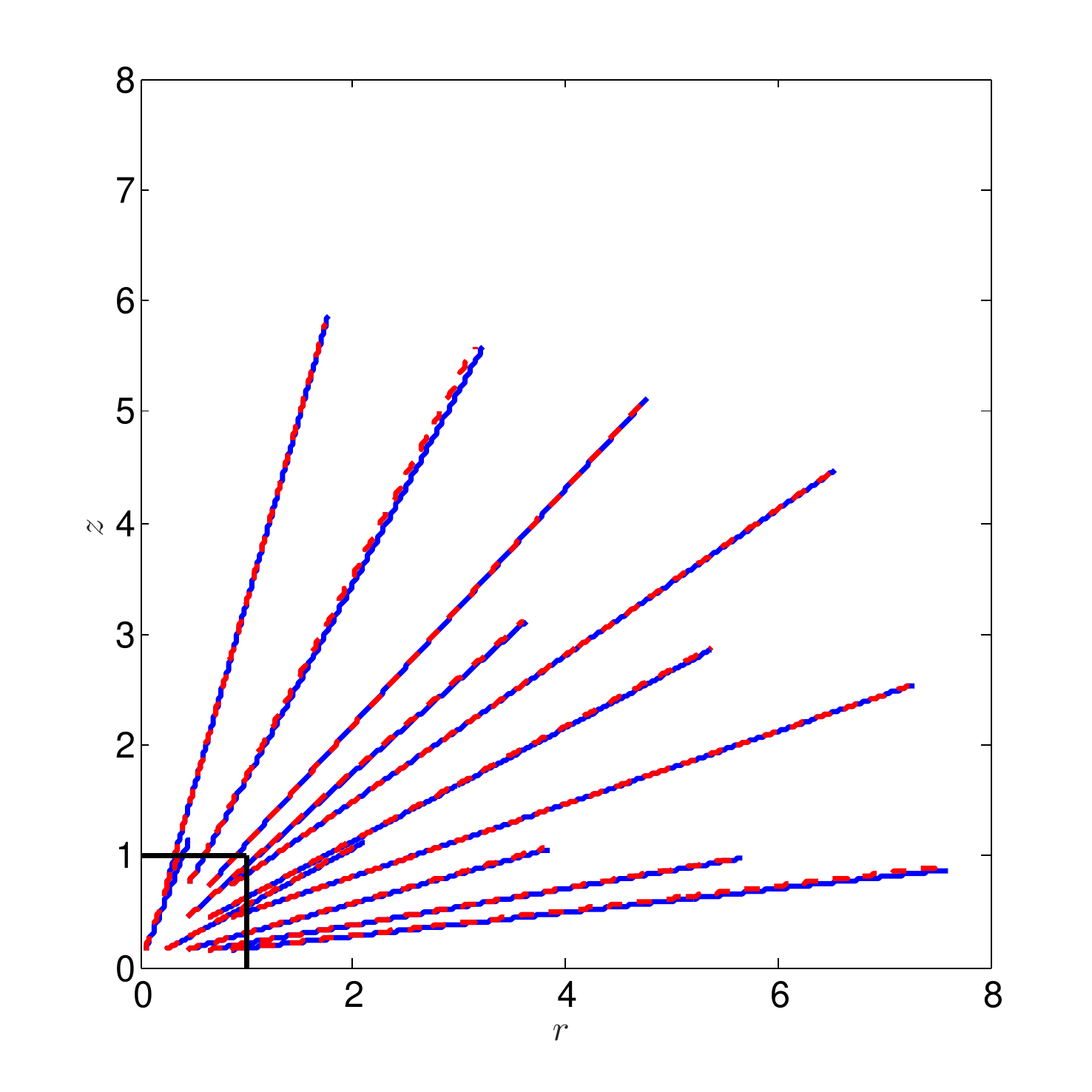}
    \caption{Same as Fig. \ref{fig:sfera_random2_4}, but for a cylinder with $H/R=1$ for $t=0 \div 10$. Results obtained with the ring method (blue lines) are compared with those obtained with the PIC method (red dotted lines).}
    \label{fig:sfera_random2_2}
\end{figure}
\begin{figure}
    \centering
    \includegraphics[width=0.65\linewidth]{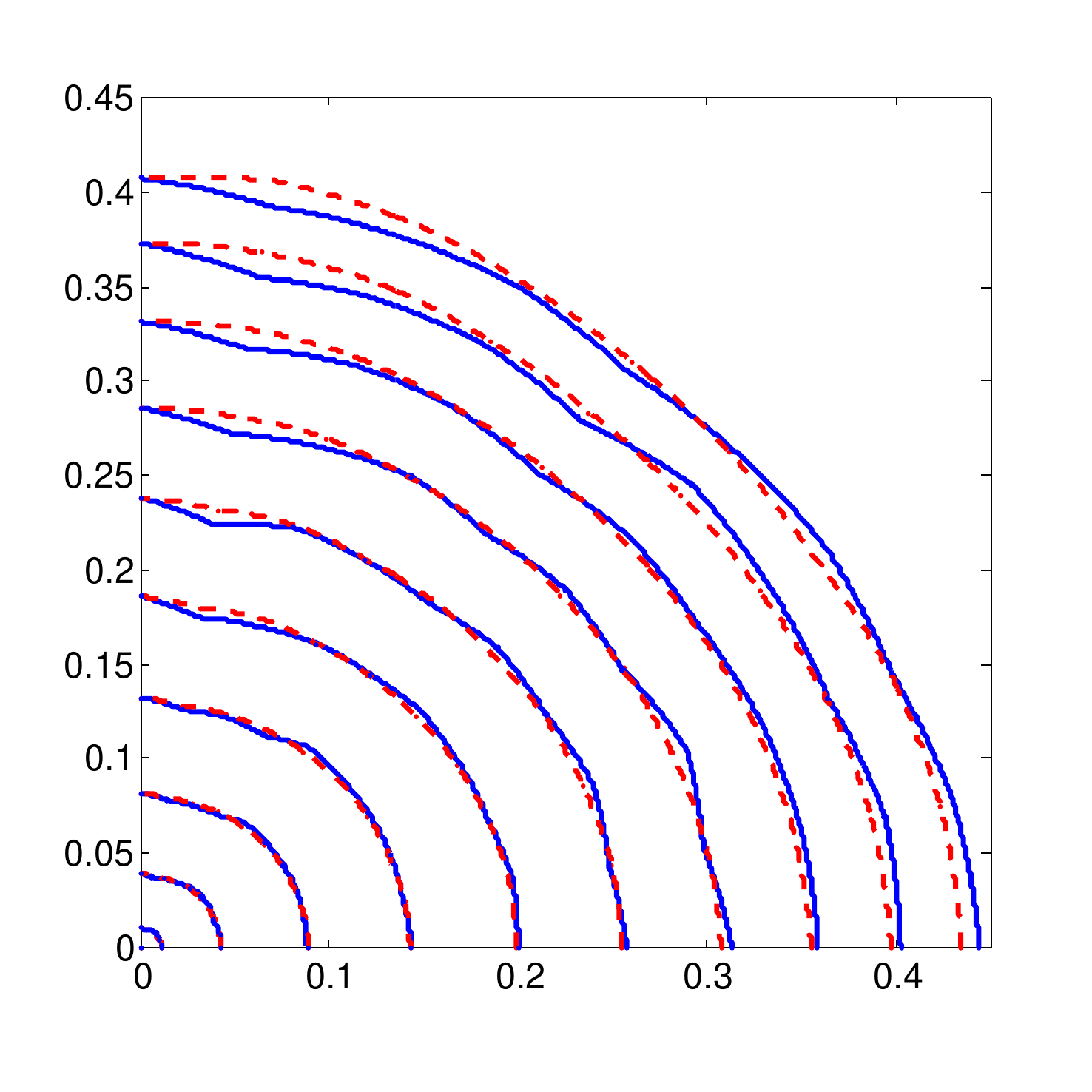}
    \caption{Same as Fig. \ref{fig:sfera_random2_3}, but for a cylinder with $H/R=1$ for $t=0 \div 10$.}
    \label{fig:sfera_random2_1}
\end{figure}

The second group of results here presented concerns the Coulomb explosion of an ion plasma having initially a cylindrical form. These are cases of practical interest, as they simulate the ion acceleration of the positive ions of a thin solid target after interaction with a ultra intense laser pulse. Two cases are considered, in which the cylinder has different aspect ratio. Figures  \ref{fig:sfera_random2_4} and \ref{fig:sfera_random2_3} show the trajectories of the ions and the
angular distribution of the kinetic energies for the first case. The same physical quantities are presented in Figs. \ref{fig:sfera_random2_2} and \ref{fig:sfera_random2_1}
for the second case. In the Figures, the results of the ring method are compared with those obtained by using a PIC code developed by the Authors \footnote{The code makes
use of an $(R,z)$ uniform grid that is expanding in order to follow the motion of the particles. Moreover, the electrostatic potential is calculated at the border of the
computational domain by summing the contributions due to all the rings; in this way, ``exact" boundary conditions are provided for the solver of the Poisson's equation.}. The agreement between the two techniques is excellent.

\section{Final considerations} \label{conclusion}
The results presented in the paper and all the tests that have been performed prove the effectiveness of the numerical technique here proposed. The interaction between computational particles is not mediated by a grid and, as shown in Sects. 2 and 3, the method can be deduced by using a Hamiltonian approach. Consequently, all the physical quantities of interest (e.g., momentum, energy and angular momentum) are conserved exactly by the method, and the only errors are due to time discretization. This properly represents an important feature of the method. When the problem has the required degree of symmetry, the methods of shells and of rings can be usefully employed in two cases: 1) to obtain results making use of a simple, easy-to-implement code; 2) to have reference results to test more complex codes, in particular when the physical region occupied by the plasma grows dramatically during the simulation. For these reasons, in the Authors' opinion the method can be regarded as a useful tool, in particular in the study of laser-plasma interaction.


\appendix

\section{Electrostatic energy of a torus with $a \ll R$}

\begin{figure}[h!]
    \centering
    \includegraphics[width=0.5\linewidth]{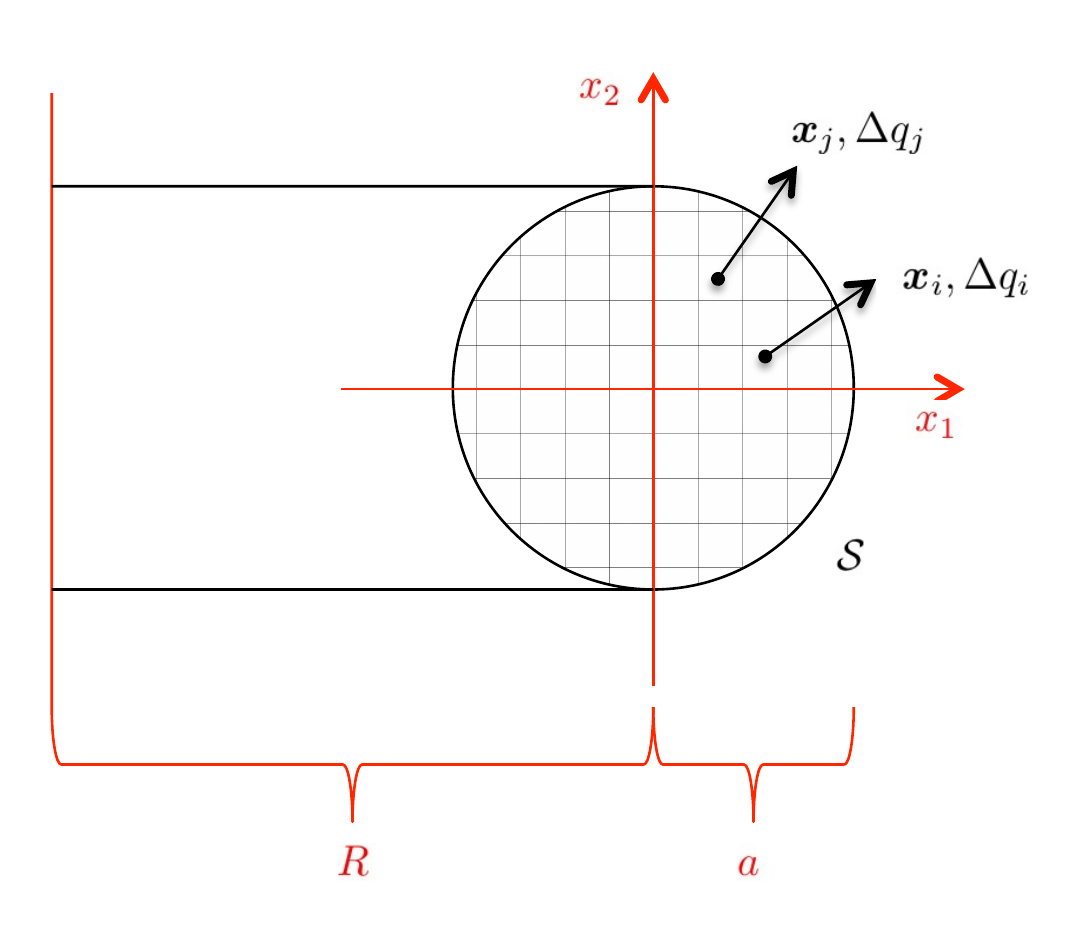}
    \caption{Cross section of a torus and coordinates employed in the calculation.}
    \label{fig:torus}
\end{figure}

With reference to Figure \ref{fig:torus}, the electrostatic energy of a torus can be calculated by dividing the cross section $S$ in a large number of subdomains.
Each of them generates an electrostatic potential that can be approximated as the one of a ring. Indicating by $\Delta q_i$ the charge of the $i$-th subdomain
and by $\varphi_{ring}(\textbf{x}_i;\textbf{x}_j)$ the potential in $\textbf{x}_i$ due to a unitary charge in $\textbf{x}_j$, the energy of the torus can be approximated by
\begin{equation}
U \simeq \frac{1}{2}\sum_{i \ne j} \Delta q_i \Delta q_j \varphi_{ring}(\textbf{x}_i,\textbf{x}_j).
\label{U_Phipozero}
\end{equation}
In the limit when the size of the subdomains tends to zero, one obtains
\begin{equation}
U=\int_{\mathscr{S}} \mathrm{d}^2x_Q \int_{\mathscr{S}} \mathrm{d}^2x_P \sigma(\textbf{x}_Q) \sigma(\textbf{x}_P) \varphi_{ring}(\textbf{x}_P,\textbf{x}_Q),
\label{U_Phipo}
\end{equation}
where $\sigma(\textbf{x})$ is the charge density for a unit cross section. If the torus is uniformly charged and if $a\ll R$, one can assume
\begin{equation}
\sigma \simeq \frac{q}{\pi a^2} = \text{Const}.
\label{U_Phipo}
\end{equation}
In order to evaluate $\varphi_{ring}(\textbf{x}_P,\textbf{x}_Q)$, the parameters $s$ and $\xi$, defined in Eq. (\ref{xi_s}),  must be evaluated. One has:
\begin{equation}
\xi=\frac{4(R+x_{1,P})(R+x_{1,Q})}{s^2}, \hspace{1cm} s=\left[(R+x_{1,P}+R+x_{1,Q})^2+(x_{2,P}-x_{2,Q})^2\right]^{1/2}.
\label{csinew}
\end{equation}
It turns out useful to introduce the quantity $\eta=R+\tfrac{x_{1,P}+x_{1,Q}}{2}$, such that $R+x_{1,P}=\eta+\tfrac{x_{1,P}-x_{1,Q}}{2}$, $R+x_{1,Q}=\eta-\tfrac{x_{1,P}-x_{1,Q}}{2}$. In this way, $\xi$ can be written as:
\begin{equation}
\xi=\frac{1-\left(\frac{x_{1,P}-x_{1,Q}}{2\eta}\right)^2}{1+\left(\frac{x_{2,P}-x_{2,Q}}{2\eta}\right)^2}\simeq 1-\left(\frac{r_{PQ}}{2R}\right)^2,
\label{csifin}
\end{equation}
with $r_{PQ}^2=\left(\textbf{x}_P-\textbf{x}_Q\right)^2$. In fact,
$\eta$ is much larger with respect to $|x_{2,P}-x_{2,Q}|\leq a$, so
the approximation $\tfrac{1}{1+\epsilon}\simeq 1-\epsilon$ can be
used; moreover, $\eta$ can be approximated by $R$. Making use of the
asymptotic behavior of $K[\xi]$ for $\xi \rightarrow 1$:
\begin{equation}
K[\xi]\underset{\xi \rightarrow 1}{\widesim[2]{}} -\frac{1}{2} \log(1-\xi)+\log 4,
\end{equation}
and assuming that $s \simeq 2R$, the following expression for $\varphi_{ring}(\textbf{x}_P,\textbf{x}_Q)$ is obtained:
\begin{equation}
\varphi_{ring}(\textbf{x}_P,\textbf{x}_Q) =-\frac{1}{\pi R} \log\left(\frac{r_{PQ}}{8R}\right).
\label{a7}
\end{equation}
Equation (\ref{a7}) can be employed in Eq. (\ref{U_Phipozero}), which can be rewritten as
\begin{equation}
U=\frac{\sigma^2}{2} \int_{\mathscr{S}} \mathrm{d}^2x_Q \varphi_{torus}(\textbf{x}_Q),
\label{ab7}
\end{equation}
being
\begin{equation}
\varphi_{torus}(\textbf{x}_Q) =-\frac{1}{\pi R} \int_{\mathscr{S}} \mathrm{d}^2x_P \log\left(\frac{r_{PQ}}{8R}\right).
\label{phitr}
\end{equation}
For $\textbf{x}_Q=0$, $\varphi_{torus}$ is readily evaluated:
\begin{equation}
\varphi_{torus}(0) =-\frac{1}{\pi R} \int_{0}^a 2\pi r dr  \log\left(\frac{r}{8R}\right)=-\frac{a^2}{R}\left[\log\left(\frac{a}{8R}\right)-\frac{1}{2}\right].
\end{equation}
To calculate $\varphi_{torus}$ for a generic $\textbf{x}_Q \in S$, one can start by noticing that $\log(r_{PQ})$ is proportional to the Green function for the two-dimensional Poisson's equation:
\begin{equation}
\nabla_Q^2\log r_{PQ} =2 \pi \delta\left(\textbf{x}_Q-\textbf{x}_P\right).
\end{equation}
So, by applying the Laplacian operator $\nabla_Q^2$ to Eq. (\ref{phitr}), one obtains
\begin{equation}
\nabla_Q^2\varphi_{torus} =-\frac{1}{\pi R} \int_S \mathrm{d}^2x_P \cdot 2 \pi \delta\left(\textbf{x}_P-\textbf{x}_Q\right)=-\frac{2}{R}.\label{ab11}
\end{equation}
Due to the symmetry of the problem, $\varphi_{torus}$ is a function
of $r_Q=|\textbf{x}_Q|$, and the Laplacian operator can be written as
\begin{math}{\nabla_Q^2=\tfrac{1}{r_Q}\tfrac{\operatorname{d}}{\operatorname{d}
{r}_Q}r_Q\tfrac{\operatorname{d}}{\operatorname{d}
{r}_Q}}\end{math}. Therefore, Eq. (\ref{ab11}) can be immediately
solved, so obtaining
\begin{equation}
\varphi_{torus}(\textbf{r}_Q) =\varphi_{torus}(0)-\frac{r_Q^2}{2R}.
\end{equation}
Finally, the energy of the torus can be calculated by using Eq. (\ref{ab7}):
\begin{equation}
U =\frac{q^2}{2\pi^2 a^4} \cdot 2 \pi \int_{0}^a r_Q \mathrm{d} r_Q \left[\varphi_{torus}(0)-\frac{r_Q^2}{2R}\right]=-\frac{q^2}{2 \pi R}\left[\log\left(\frac{a}{8R}\right)-\frac{1}{4}\right].
\label{ab13}
\end{equation}
Formula (\ref{ab13}) is very accurate for $a \ll R$. If compared
with the value of $U$ obtained from numerical integration of Eq.
(\ref{U_Phipozero}), the relative error is less than 0.5$\%$ for
$a/R<0.2$. A similar formula (without the term -1/4) has been
deduced in a concise, brilliant way in \cite{landau8} by using the
technique of asymptotic matching.

\label{Appendix2}

\newpage
\bibliographystyle{elsarticle-num}
\bibliography{draft}






\end{document}